\documentclass[preprint,12pt]{elsarticle}
\usepackage{float}
\usepackage{psfrag}
\usepackage{amsmath}
\usepackage{amsfonts}
\usepackage{graphicx}
\usepackage{color}
\usepackage[utf8]{inputenc}
\usepackage{lscape}
\usepackage{multicol}
\usepackage{srcltx}
\usepackage{overpic}
\usepackage{titletoc}
\usepackage{multirow}
\usepackage{tikz}
\usepackage{enumitem}
\usepackage[papersize={216mm,279mm},tmargin=20mm,bmargin=30mm,lmargin=30mm,rmargin=30mm]{geometry}

\journal{Rev. Mex. Phys. E}

\begin{document}
\markboth{D. Sanjin\'es, J. Velasco and E. Mamani}{THE KEPLER PROBLEM ON THE LATTICE}

\title{The Kepler problem on the lattice
\author{D. Sanjin\'es, E. Mamani}
\address{Carrera de F\'isica, Universidad Mayor de San Andr\'es, c. 27 Cota - Cota, casilla de correos 8635, La Paz, Bolivia.\\email: diegosanjinescastedo@gmail.com; evaristomamanicarlo@gmail.com}
\author{J. Velasco}
\address{Departamento de F\'isica, Universidad T\'ecnica de Oruro, Oruro, Bolivia.\\email: jvelascov110@gmail.com}}

\begin{abstract}
We study the motion of a particle in a 3-dimensional lattice in the presence of a potential $-V_1/r$, but we demonstrate semiclassicaly that the trajectories will always remain in a plane which can be taken as a rectangular lattice. The Hamiltonian model for this problem is the conservative tight-binding one with lattice constants $a, b$ and hopping elements $A, B$ in the $XY$ axes, respectively. We use the semiclassical and quantum formalisms; for the latter we apply the pseudo-spectral algorithm to integrate the Schr\"{o}dinger equation. Since the lattice discrete subspace is not isotropic, the angular momentum is not conserved, which has interesting consequences as chaotic trajectories and precession trajectories, similar to the astronomical precession trajectories due to non-central gravitational forces, notably, the non-relativistic Mercury's perihelion precession. Although the elements of the mass tensor are naturally different in a rectangular lattice, these can be chosen to be still different in the continuum, which permits to study the motion with kinetic energies $p_i^2/2m_i$ ($i=x,y$). We calculate also the contour plots of an initial Gaussian wavepacket as it moves in the lattice and we propose an ``intrinsec angular momentum" $S$ associated to its asymmetrical deformation, such that the quantum and semiclassical angular momenta, $L_q, L_c$, respectively, could be related as $L_q=L_c+\alpha S$.
\end{abstract}

\begin{keyword}
Kepler problem; tight-binding dynamics; semiclassical model; pseudo- spectral method; physics education.
\end{keyword}

\maketitle

\newpage

\section{Introduction}

The Kepler problem on the lattice can be traced back to the anisotropic Kepler problem about the description of an electron dynamics in the presence of the Coulomb field of a donor impurity in a Silicon or Germanium semiconductors \cite{Luttinger1954}. Some authors have revived this issue either in the theoretical or the experimental contexts, for example, because of more efficient numerical algorithms and simulation applications, or new and more precise experimental techniques. We may mention a few authors in the theoretical context: (i) Boris et al. \cite{Boris1993} use quantum analytical and numerical methods to describe the wavepaket evolution within elliptical and hyperbolic orbits. (ii) Bai and Zheng \cite{Bai2002} invoke classical and quantum mechanics methods for studying the Kepler problem with a kind of weak anisotropy of the mass tensor. (iii) Petrova \cite{Petrova2016} set about solving exactly the full energy spectrum (via Green's function methods) of a particle hopping on a Bethe lattice under a Coulomb potential. (iv) Cort\'es \cite{Cortes2008} has particularly caught our attention because he obtains a chaotic particle's trajectory by defining an anisotropic mass tensor, a result that we could also obtain by defining the Kepler problem on a rectangular lattice, such that in the continuum limit the lattice disappears but an anisotropic mass tensor remains.

The main motivation for our work presented in this article has an educational purpose: the Kepler problem is a common topic taught at the undergraduate level (a usual reference text is Symon \cite{Symon1953}) where the conic-section trajectories are deduced from the conservation of energy and angular momentum. As pointed out above \cite{Cortes2008}, the definition of an anisotropic mass tensor, and consequently the non-conservation of angular momentum, leads to a simple system where chaos was rigorously proved, becoming therefore a common educational model to teach this subject. Within this same purpose, we realize that the introduction of a ``fake lattice" in the space can be used as a model to justify the mass asymmetry, not as a consequence of some particle's property, but as a property of space itself lacking rotational symmetry. Moreover, two important and rather different subjects in the undergraduate physics education, namely, the astronomical Kepler problem (studied in classical mechanics \cite{Symon1953}) and the tight-binding dynamics (studied in solid state by quantum and semiclassical methods \cite{Mermin1976}) converge in the ``Kepler problem on the lattice". This, in turn, is the ocassion to draw other subjects in physics education, as for example: higher order statistical moments associated to random distributions \cite{Walpole1980} and their relation to physical properties of the quantum probability density (as suggested in this article); analytical and numerical methods to solve the Schr\"{o}dinger equation \cite{Liboff1980} \cite{DeVries1994}.

Our work in this article is organized as follows: In Sections 2 and 3 we set about developing the semiclassical and quantum formalisms, respectively, at the common level of the usual textbooks for the undergraduate education in physics, for example, Symon \cite{Symon1953} and Liboff \cite{Liboff1980}. In Section 4 we present the central quantum results of our work and compare them to previous semiclassical results. In Section 5 we give a list of the most relevant results of our work.

\section{Semiclassical model for the rectangular \text{lattice}}

\begin{figure}[H]
\centering
\includegraphics[scale=0.30]{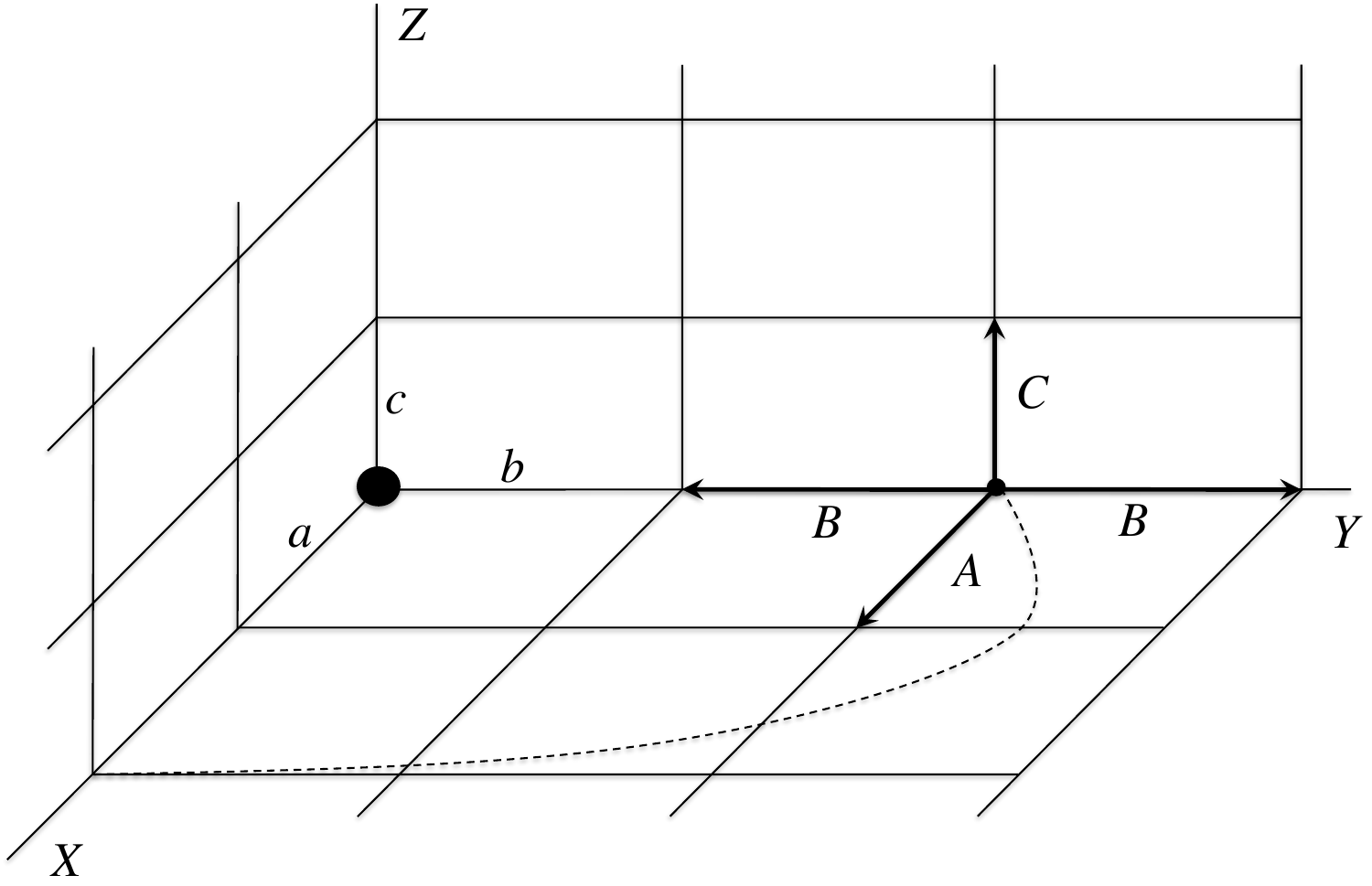}
\caption{Case of the Kepler problem on the lattice where the impurity is located at the origin of the orthorhombic lattice; the electron is initially located at the position $\textbf{r}_0=(0,2b,0)$ subject to the impurity Coulomb interaction and to the 1st-neighbor interactions with energies (hopping elements) $A, B, C$ along the $X, Y, Z$ axes, respectively. If the initial velocity lies on the $XY$ plane, the particles's trajectory will also remain in the plane (dotted line).}
\label{fig1}
\end{figure}

The tight-binding Hamiltonian function for an electron moving by the hopping mechanism in a 3-dimensional orthorhombic lattice with constants $a, b, c$ along the $X, Y, Z$ axes and hopping elements $A, B, C$, respectively, and in the presence of a donor impurity potential located at the origin, is:
\begin{align}
H(\textbf{r},\textbf{k})=&2A(1-\cos{ak_x})+2B(1-\cos{bk_y})\nonumber\\&+2C(1-\cos{ck_z})-V_1/\sqrt{x^2+y^2+z^2}.\label{1}
\end{align}This function has been constructed, in accordance with the semiclassical method \cite{Mermin1976}, from the spectrum of the tight-binding Hamiltonian operator for a particle hopping in a periodic potential $V(\textbf{r}+\textbf{R})=V(\textbf{r})$ with $\textbf{R}=(na,mb,qc)$ ($n,m,q$ integers) and $\hbar\textbf{k}=\hbar (k_x,k_y,k_z)$ the position and quasimomentum vectors, respectively, in 3D orthorhombic Bravais lattices. In the limit of the effective mass approximation wherein $m_x=\lim_{A\rightarrow \infty, a\rightarrow 0} \hbar^2/(2Aa^2)$, the terms $\hbar k_x$ and $\hbar^2 k_x^2/(2 m_x)$ are the usual momentum and energy, respectively, of a free particle along the $X$ direction. The term $V_1$ in (\ref{1}) is the coefficient (in units of \textit{joule-meter}) of the Coulomb potential energy due to an \textit{external} point source (not intrinsic to the lattice) which is fixed somewhere \textit{upon} the lattice and referred to in this article as \textit{the impurity}. Fig.\;\ref{fig1} illustrates a case of the Kepler problem on the lattice described above.

The corresponding Hamilton equations of motion (with $\textbf{p}=\hbar \textbf{k}$) are
\begin{align}
\dot{x}=\frac{\partial H}{\partial p_{x}}=(2Aa/\hbar)\sin{ak_{x}},\label{2}\\
\dot{y}=\frac{\partial H}{\partial p_{y}}=(2Bb/\hbar)\sin{bk_{y}},\label{3}\\
\dot{z}=\frac{\partial H}{\partial p_{z}}=(2Cc/\hbar)\sin{ck_{z}},\label{4}\\
\hbar \dot{k}_{x}=-\frac{\partial H}{\partial x}=V_1 x (x^2+y^2+z^2)^{-3/2},\label{5}\\
\hbar \dot{k}_{y}=-\frac{\partial H}{\partial y}=V_1 y (x^2+y^2+z^2)^{-3/2},\label{6}\\
\hbar \dot{k}_{z}=-\frac{\partial H}{\partial z}=V_1 z (x^2+y^2+z^2)^{-3/2}.\label{7}
\end{align}These equations along with their corresponding initial conditions at $t=0$ for the position $\textbf{r}(0)$ and velocity $\textbf{v}(0)=\dot{\textbf{r}}(0)$, constitute the semiclassical model for the tight-binding dynamics of a hopping particle in a lattice.
By choosing the initial conditions such that the particle moves in the $XY$ plane and is not located at the origin: $z(0)=\dot{z}(0)=0$, $x^2(0)+y^2(0)\neq 0$, we deduce from (\ref{7}) that $\dot{k}_z(0)=0$ and from (\ref{4}) that $k_z(0)=0$; therefore $\ddot{z}(0)=2Cc^2\dot{k}_z(0)\cos(ck(0))=0$, i.e., the particle will continue moving in the $XY$ plane in the abscence of any perpendicular acceleration that could drag the particle out of the plane. Henceforth, we will consider the dynamics of the electron in a rectangular lattice described by Eqs. (\ref{2}), (\ref{3}), (\ref{5}) and (\ref{6}) above (with $z=0$). Besides, and for simplicity of notation, we will take a unitary value for the angular Planck constant, $\hbar=1$. This would imply, of course, that the momentum and energy should be expressed in units of 1/\textit{meter} and 1/\textit{second}, respectively, but the election of $\hbar=1$ in this article does not bear any specific physical meaning, since the true numerical value of $\hbar$ would be restored before interpreting physical measurements.

From the definition of angular momentum, \text{$\textbf{L}=\textbf{r}\times\textbf{p}$}, we have the component of $\textbf{L}$ in the rectangular lattice as $L_z=xk_y-yk_x$. Its corresponding time-variation is $\dot{L}_z=2\left(aAk_y\sin{ak_x}-bBk_x\sin{bk_y}\right)$, which is different from zero for an arbitrary instant. This means that in the rectangular lattice the angular momentum $L_z$ is not conserved in the presence of the point impurity located at the origin with a corresponding Coulomb singular potential $-V_1/\sqrt{x^2+y^2}$, equivalent to the gravitational potential of a point mass in the two-body Kepler problem. However, for these problems we know from our background in classical mechanics that the angular momentum is conserved (along with the total energy) because the Coulomb and gravitational forces are central and therefore the torque is zero. What is assumed in these cases is that the space is rotationally symmetric (isotropic); that is not valid for the rectangular lattice. The procedure to achieve an isotropic space from the rectangular lattice is, first, to make a square lattice out of a rectangular one by making $a=b$ and $A=B$ in the Hamiltonian function $H$ in (\ref{1}), and second, to take the continuum limit as $a \rightarrow 0$ while $A \rightarrow \infty$, such that $1/m=2Aa^2$ is defined as the ``effective mass". In fact, since the cosine approximates as $\cos (a k_x)\cong 1-a^2 k_x^2/2$ for small arguments, we have that
\begin{align}
H\left(x,y;k_{x},k_{y}\right) \rightarrow (k_x^2+k_y^2)/2m-V_1/\sqrt{x^2+y^2},\label{8}
\end{align}giving thus the usual Hamiltonian function for the 2-dimensional Kepler problem. We can see that in this case $\dot{L}_z \rightarrow 0$, i.e., the angular momentum is conserved, as expected. Interestingly, if the continuum limit is obtained from the rectangular lattice by maintaining $a \neq b$ and $A \neq B$, we obtain
\begin{align}
H\left(x,y;k_{x},k_{y}\right) \rightarrow k_x^2/2m_x+k_y^2/2m_y-V_1/\sqrt{x^2+y^2},\label{9}
\end{align}along with $\dot{L}_z \neq 0$, which describes the classical anisotropic Kepler problem reported by Cort\'es \cite{Cortes2008} where the angular momentum is not conserved due to the difference of effective masses $m_x, m_y$ in the $XY$ axes, respectively. As we can see, in this case it is suggested that the ``anisotropy parameter" defined as $\xi \equiv m_x/m_y$ \cite{Cortes2008} has its origin in the quotient $m_x/m_y=Bb^2/(Aa^2)$ of a \textit{subyacent} rectangular lattice in the continuum limit. In order to reduce the free parameters in that quotient we can invoke an expression reported by Wolf \cite{Wolf2014} wherein the hopping element is related to the separation by a lattice constant between 1st. neighbor sites: $A=2(1+a/a_0)\exp(-a/a_0)$ and $B=2(1+b/a_0)\exp(-b/a_0)$, where $a_0$ is the Bohr radius.

\begin{figure}[H]
\centering
\includegraphics*[scale=0.45]{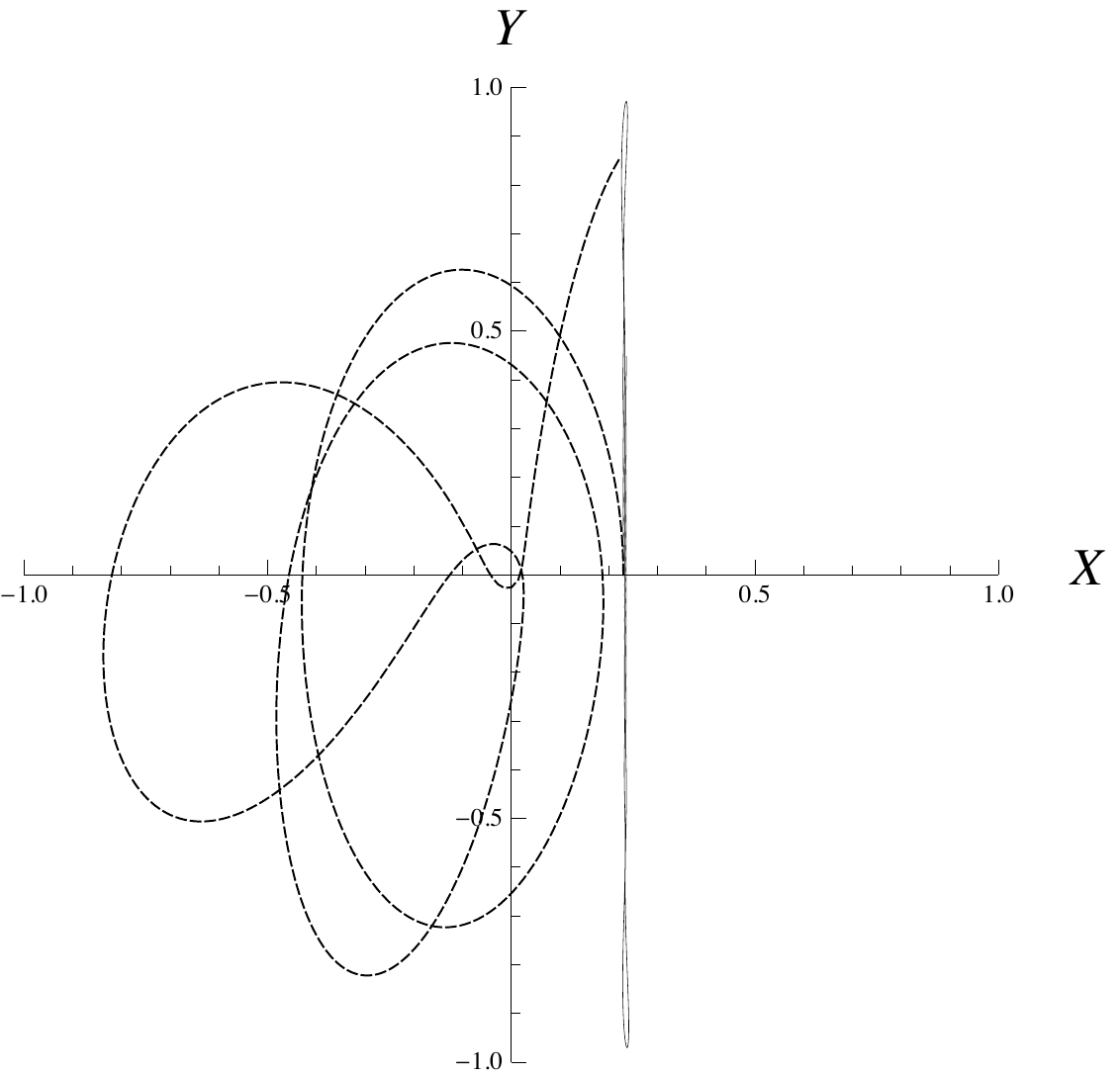}
\caption{Trajectory of the particle in the $XY$ plane of a rectangular lattice with a ratio $a/b \cong 20$ of the lattice constants (solid line). The initial conditions are $x(0)=0.23a, y(0)=0, k_x(0)=0$ and $k_y(0)=5.33/b$. The resulting motion yields an oscillatory trayectory parallel to the $Y$ axis. In the continuum limit $a,b \rightarrow 0$ (maintaining $a/b \cong 20$) and $A,B\rightarrow \infty$ corresponding to the Hamiltonian (\ref{9}), and $k_y(0)=400.68/b$, the trajectory becomes the dashed one. This result reproduces the one reported by Cort\'es \cite{Cortes2008} in the continuum, who introduces different masses $m_x, m_y$ in the $XY$ axes. Our result show that the rectangular lattice approach to the continuum renders a satisfactory explanation for the origin of the particle's mass anisotropy.}
\label{fig2}
\end{figure}

In fact, following Cort\'es \cite{Cortes2008}, we take $\xi=2.94$ as the anisotropy parameter and solve $\xi=Bb^2/(Aa^2)$ with $a/a_0=9.5$ which yields the solution $b/b_0=0.477$. The corresponding trajectory in the $XY$ plane is shown in Fig.\;\ref{fig2} (dashed line); it agrees quite well with the one reported by Cort\'es. Moreover, we can preserve the ratio $a/b \cong 20$ for a finite lattice without taking the continuum limit $a, b\rightarrow 0$; the result for the same initial conditions is also shown in Fig.\;\ref{fig2} (solid line). We can see a radical different trayectory characterized by an almost 1-dimensional oscillation along the $Y$ axis and $x=x(0)=0.23$ constant. The explanation of this phenomenon is the following: in the finite lattice, the ratio $a/b \cong 20$ means that the  probability of the particle's hopping to neighbor sites along the $X$ axis is neglible as compared to the sites along the $Y$ axis, resulting thus in an effective  motion due to the $Y$ component force of the Coulombian attraction; this component is a restoring force with a magnitude $V_1 (x(0)^2+y^2)^{-3/2} y$. For the values of the simulation in Fig.\;\ref{fig2}: $x(0)=0.23$ and $-1<y<1$, the resulting motion is oscillatory; as the condition $x(0)\gg y$ is approached, the restoring $Y$ force is approximated as a linear one, resulting thus in a simple harmonic motion. In the continuum limit the hopping along the $X$ axis sites is no longer neglible and the particle follows a 2-dimensional trajectory.

\begin{figure}[H]
\centering
\includegraphics*[scale=0.42]{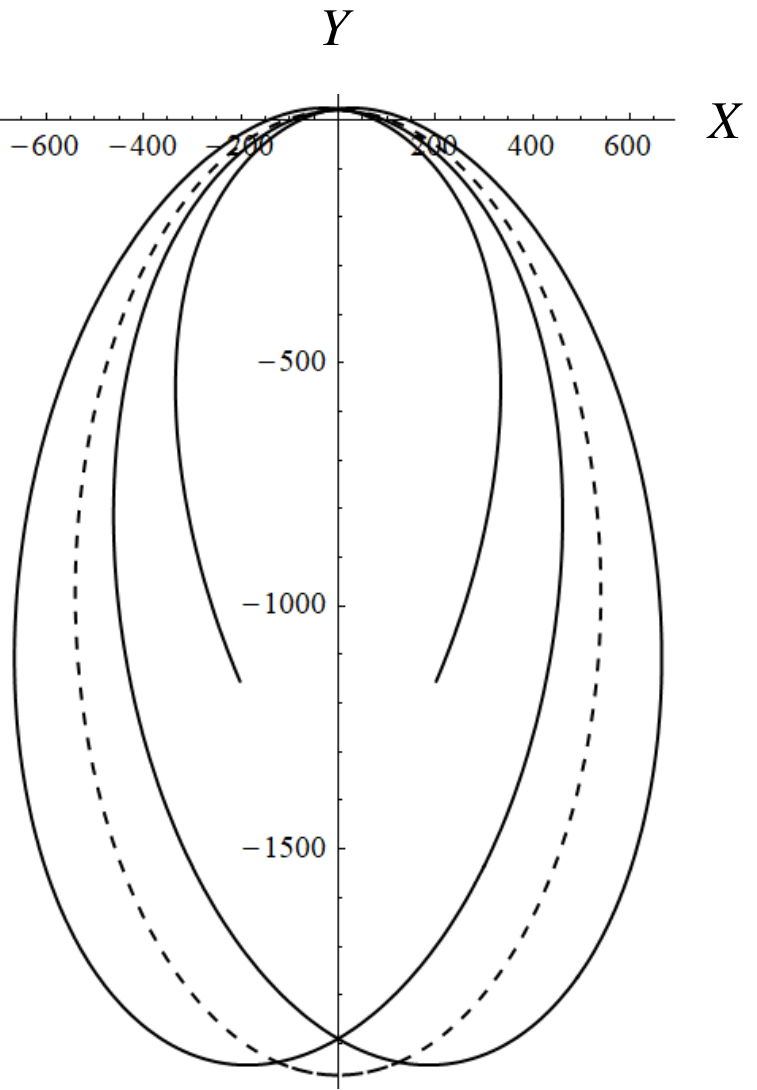}
\caption{``Precession of the perihelion" trajectory of a particle in the $XY$ plane of a square lattice with adimensional parameters $a=1$, $A=125$, $V_1=20000$ in the 2-dimensional Hamiltonian (\ref{1}) and conditions at $t_0 \cong 0.17\;T$ in the ``perihelion" position: $(x_0,y_0)=(0,20)$, $(k_{x0},k_{y0})=(-1,0)$. The impurity is located at $(0,-140)$. In the continuum limit the precession effect disappears and the quasi-elliptical orbits collapse into a unique elliptical orbit (dashed line) with period $T$.}
\label{fig3}
\end{figure}

\begin{figure}[H]
\centering
\includegraphics*[scale=0.48]{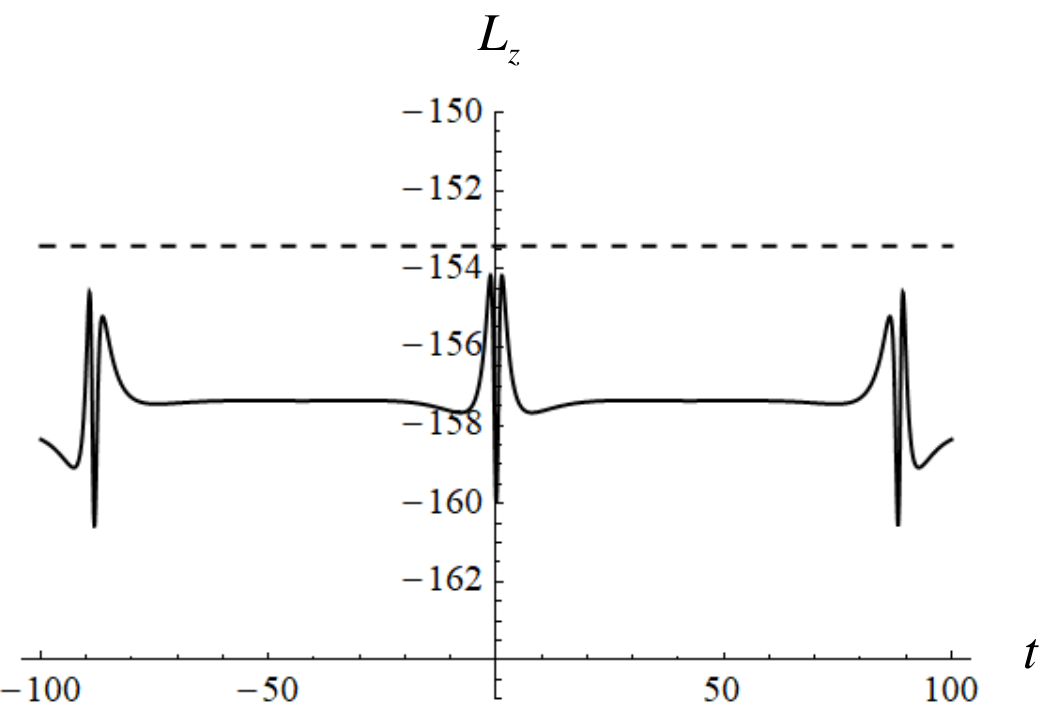}
\caption{Angular momentum $L_z(t)=xk_y-yk_x$ corresponding to the cases in Fig.\;\ref{fig3}. $L_z$ changes abruptly at the perihelion positions of the precession orbits (solid line) while in the continuum limit $L_z$ tends to a constant value $L_z^{\textrm{cont}}$ (dashed line). The difference of $L_z^{\textrm{cont}}$ and the plateau value of $L_z \cong L_z^{\textrm{aph}}$ near the aphelion position is due to the difference of values for $k_x(0)$ at the position $(0,y_0)$ that guarantee the conservation of energy.}
\label{fig4}
\end{figure}

Interestingly, if the rectangular lattice is transformed into a square one ($a=b, A=B$) and we take the continuum limit, the resulting trajectory is an ellipse (Fig.\;\ref{fig3}, dashed line), as expected in an isotropic continuum space with a central force field, while in the lattice the trajectory is a succesion of quasi-elliptical orbits (Fig.\;\ref{fig3}, solid line) giving rise to the familiar trajectory showing the precession of the perihelion due to non-central forces in Newtonian celestial mechanics \cite{Symon1953}. The behaviour of the angular momentum $L_z=xk_y-yk_x$ corresponding to the cases in Fig.\;\ref{fig3} is shown in Fig.\;\ref{fig4}. The solid line shows the abrupt change of $L_z$ in the perihelion, as expected, while in the rest of the quasi-elliptical trajectory passing through the aphelion $L_z \cong L_z^{\textrm{aph}}$ remains almost constant and equal to their values for other orbits. In the continuum limit, $L_z$ tends to a constant value $L_z^{\textrm{cont}}$ (dashed line), again, as expected. The evident difference between $L_z^{\textrm{aph}}$ and $L_z^{\textrm{cont}}$ is due to the fact that in order to maintain constant the value of the energy $E=2A(1-\cos ak_x(0))-V_1/y_0$, the initial condition for $k_x(0)$ at $(0,y_0)$ should change according to

\begin{align}
k_x(0)&=\frac{1}{a}\cos^{-1}\left[1-\frac{a^2}{2}(k^{\textrm{cont}}_x(0))^2\right],\label{9a}
\end{align}where $k_x(0) \rightarrow k^{\textrm{cont}}_x(0)$ as $a \rightarrow 0$.

The results reported in this section concern the application of the semiclassical model to the Kepler problem on the lattice, taking as the starting point the Hamiltonian function (\ref{1}) which was constructed from the tight-binding energy of the lattice and the Coulomb potential energy due to the impurity. Since such a tight-binding energy is deduced within the quantum formalism (see, for example, ch.10 in \cite{Mermin1976}), the semiclassical results can only be considered as approximate, being the quantum formalism the exact theoretical framework to study the Kepler problem on the lattice. It is known that the semiclassical position $\textbf{r}$ and momentum $\hbar \textbf{k}$ defined in (\ref{2})--(\ref{7}) correspond to their quantum analogues as the mean values of the position $\langle\Psi|\textbf{r}|\Psi\rangle$ and momentum $\langle\Psi|-i\nabla|\Psi\rangle$ operators for a normalized quantum state $\Psi(\textbf{r},t)$. For such an approximation to hold, the semiclassical model is restricted by the conditions that the quantum wavepacket width should be much larger than the lattice constants $a, b, c$ in (\ref{1}) and much smaller than the length of an appreciable variation of the external Coulomb field. Besides, it is required that the energy $V_1/r$ of the external field in (\ref{1}) should not be too large so that the wavepacket's energy remains within a single band (see, for example, ch.12 in \cite{Mermin1976}). It is also known that the wavepacket spreads as it propagates in the lattice, being in general such spreading non-symmetrical with respect to the wavepacket's center of mass. Therefore, if a such a non-symmetrical spreading could be related to a kind of ``internal degree of freedom" of the wavepacket, then it would be possible to establish a significant relation between other dynamical variables than the position and momentum. Those dynamical variables are the semiclassical angular momentum and its quantum equivalent. This is done in the next two sections.

\section{Quantum formalism in the lattice}

The formal solution of the time-dependent Schr\"{o}dinger equation $i \partial \Psi/ \partial t=\widehat{H} \Psi$ (recall that $\hbar=1$) for a general time-independent one-particle Hamiltonian operator $\widehat{H}$, is given as \cite{Liboff1980}

\begin{align}
\Psi(\textbf{r},t)&=e^{-i\widehat{H}t} \Psi(\textbf{r},0),\label{10}\\
\Psi(\textbf{r},t+\Delta t)&=e^{-i\widehat{H}\Delta t} \Psi(\textbf{r},t)\\
&=e^{-i\widehat{V}\Delta t/2} e^{-i\widehat{T}\Delta t} e^{-i\widehat{V}\Delta t/2}\Psi(\textbf{r},t)+O(\Delta t ^3),\label{11}
\end{align}where the Hamiltonian was separated, within the single-band approximation \cite{Mattis1986}, into a kinetic energy operator $\widehat{T}(\textbf{p})$ (with $\textbf{p}=-i\nabla$ in the position representation) and a external potential energy operator $\widehat{V}(\textbf{r})$; the periodicity of the lattice is incorporated within $\widehat{T}(\textbf{p})$ while the potential energy corresponds to externally applied forces not intrinsic to the lattice (such as the electric field). The further splitting of the Hamiltonian as $\widehat{H}=\widehat{V}/2+\widehat{T}+\widehat{V}/2$ follows from the application of Baker-Campbell-Hausdorff formula \cite{Swanson1992} for the non-commuting operators $\widehat{T}$ and $\widehat{V}$, giving thus an approximate time-iterative formula whose integration yields the result in (\ref{10}) within a quite good approximation for a finite $\Delta t$.

The next step is to transform the algorithm (\ref{11}) into a practical useful tool for integrating the Schr\"{o}dinger equation by expressing (\ref{11}) in terms of Fourier ($\mathcal{F}$) and inverse Fourier ($\mathcal{F}^{-1}$) transforms as

\begin{align}
\Psi(\textbf{r},t+\Delta t)=&e^{-iV(\textbf{r})\Delta t/2} \mathcal{F}^{-1}\left[e^{-iT(\textbf{k})\Delta t}\right.\nonumber\\
&\left.\times\mathcal{F}\left[e^{-iV(\textbf{r})\Delta t/2}\Psi(\textbf{r},t)\right]\right]+O(\Delta t^3).\label{11a}
\end{align}In this expression, the operator $\widehat{T}(\textbf{p})$ is transformed into the function $T(\textbf{p})$ (with $\textbf{p}=\textbf{k}$) in the reciprocal space where the inverse Fourier transform ($\mathcal{F}^{-1}$) acts upon \cite{DeVries1994}. The operator $\widehat{V}(\textbf{r})$ is just the function $V(\textbf{r})$ in the direct space where the Fourier transform ($\mathcal{F}$) acts upon. Specifically, for the Kepler problem on the lattice defined by the Hamiltonian function $H$ in (\ref{1}), $T(\textbf{k})=2A(1-\cos ak_x)+2B(1-\cos bk_y)$ and $V(\textbf{r})=-V_1 (x^2+y^2)^{-1/2}$, already applied within the $XY$ plane, as explained below (\ref{7}).

To complete the description of the quantum formalism in the lattice, we use the representation of the state $\Psi(\textbf{r},t)$ in the basis of the complete set of orthonormal Wannier functions $\Phi(\textbf{r}-\textbf{R})$ (the Wannier basis) as $\Psi(\textbf{r},t)=\sum_{n,m} C_{nm} (t)\Phi(\textbf{r}-\textbf{R})$, where these functions are localized about the lattice site $\textbf{R}=(na,mb)$ ($n,m$ integers) with an extent of the order of the lattice constants $a$ and $b$ in the rectangular lattice [19]. Therefore, the algorithm (\ref{11a}) is to be applied to the coefficients $C_{nm}(t)$ (which represent the state $\Psi(\textbf{r},t)$) to calculate the $p$-th order statistical moments.

The case of $p=0$ corresponds to the normalization condition of the quantum wavepacket, $\sum_n |C_{nm}|^2=1$. For $p=1$ we have the mean values of the position along the $X$ and $Y$ axes: $z_x=\sum_{n,m}an|C_{nm}|^2$ and $z_y=\sum_{n,m}bm|C_{nm}|^2$, respectively. The 2nd moment about the mean (along the $X$ axis) is the variance: $\sigma_x^2=\sum_{n,m}(an-z_x)^2|C_{nm}|^2$, whose square root is the standard deviation $\sigma_x$. Interestingly, the mean values of the position $z_x, z_y$ correspond to their classical analogues $x, y$ in the semiclassical model under the specific physical conditions referred to at the end of Section 2. The variances $\sigma_x^2, \sigma_y^2$ represent the dispersion of the quantum wavepackets relative to their mean positions but these variances do not have a dynamical variable counterpart. The 3th moments will be considered in this work as dynamical significant in the sense we will explore below. They will be defined here as $s_x^3\equiv\sum_{n,m}(an-z_x)^3|C_{nm}|^2$ and the corresponding expression for $s_y^3$, instead of the so-called ``skewness" which is defined as $s_x^3/\sigma_x^3$ and $s_y^3/\sigma_y^3$ \cite{Walpole1980}. The skewness is a measure of the asymmetry of the wavepacket distribution about the mean. Notice that while the skewness has no physical dimensions, $s_x, s_y$ have dimensions of length, and this will suit our purposes hereafter.

Finally, we calculate the mean value $\langle\textbf{L}\rangle$ of the angular momentum operator $\textbf{L}=\textbf{r} \times \textbf{p}$ ($\textbf{p}=-i\nabla$) in the quantum state approximated as  $\phi(\textbf{r})=D \exp\left({i\int^\textbf{r} \textbf{k}(\textbf{r}') \cdot d\textbf{r}'}\right)$ according to the zeroth-order term of the WKB expansion \cite{Liboff1980}. The result is $\langle\textbf{L}\rangle \equiv L_q=|\textbf{z}\times \textbf{k}|=z_x k_y-z_y k_x$, where $z_x, z_y$ are the mean values of the $x, y$ coordinates calculated in the lattice as in (\ref{12}) and (\ref{13}) below.

\section{Quantum numerical simulations}

In this Section we apply the algorithm (\ref{11a}) in terms of direct and inverse Fourier transforms to the coefficients $C_{nm}(t)$ which represent the quantum state $\Psi(\textbf{r},t)$. The recursive application of (\ref{11a}) over a time interval is thus equivalent to solve numerically the Schr\"{o}dinger equation. Then, with the values of $C_{nm}(t)$, we calculate specifically the first order and third order statistical moments as defined in Section 3, which correspond to the wavepaket's center of mass and ``skewness", respectively.

For 3-dimensional quantum simulations in a cubic lattice ($a=b=c$) we use

\begin{align}
z_x&=\sum_{n,m,p}an|C_{nmp}|^2,\label{12}
\end{align}
\begin{align}
z_y&=\sum_{n,m,p}am|C_{nmp}|^2,\label{13}
\end{align}
\begin{align}
z_z&=\sum_{n,m,p}ap|C_{nmp}|^2,\label{14}
\end{align}for the mean-value positions of the quantum wavepacket.

\begin{figure}[H]
\centering
\includegraphics*[scale=0.60]{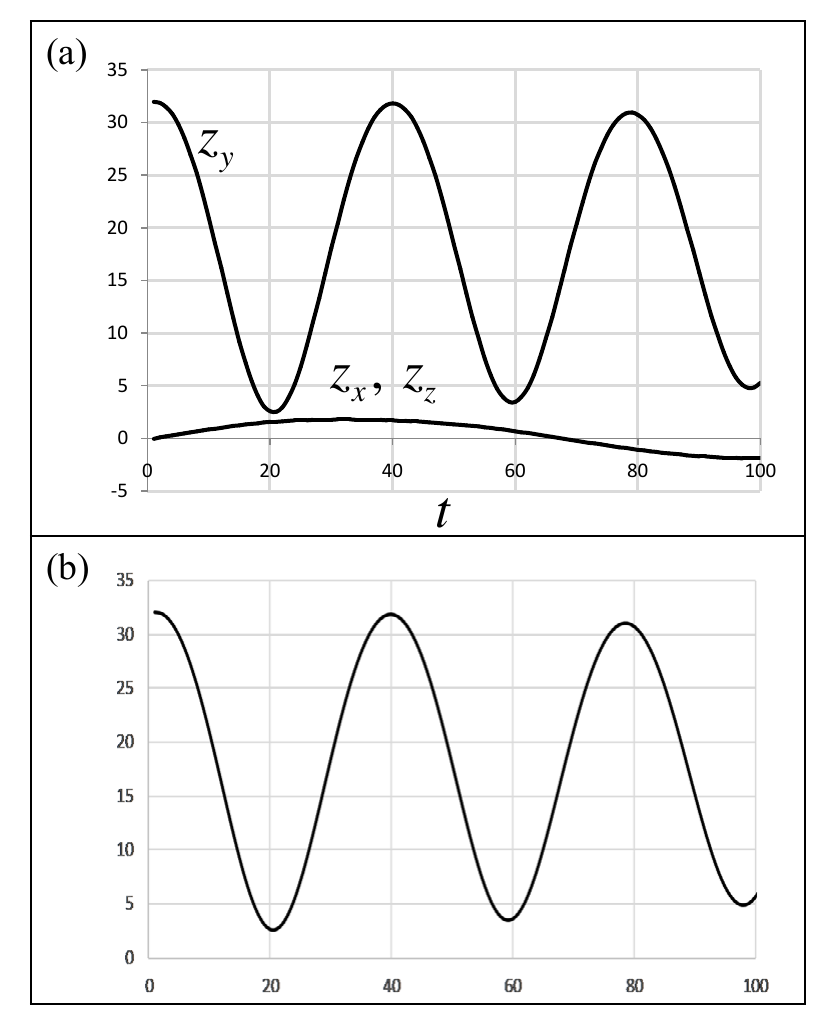}
\caption{Quasi-Bloch oscillations: values of $z_y(t)$ for the initial Gaussian wavepacket with conditions $(x_0,y_0)=(0,32a)$ and $(k_{x0},k_{y0})=(0,0)$; the donor impurity acting as center of the Coulomb attraction is located at $(0,-120a)$ and the intensity of the Coulomb potential in (\ref{1}) is $V_1=307200$ (adimensional units). Figure (a) corresponds to the 3-dimensional simulation with negligible and equal values of $z_x$ and $z_z$; figure (b) corresponds to the 2-dimensional simulation in the $XY$ plane where $z_x$ (not shown) is the same as in case (a).}
\label{fig5}
\end{figure}

\begin{figure}[H]
\centering
\includegraphics*[scale=0.40]{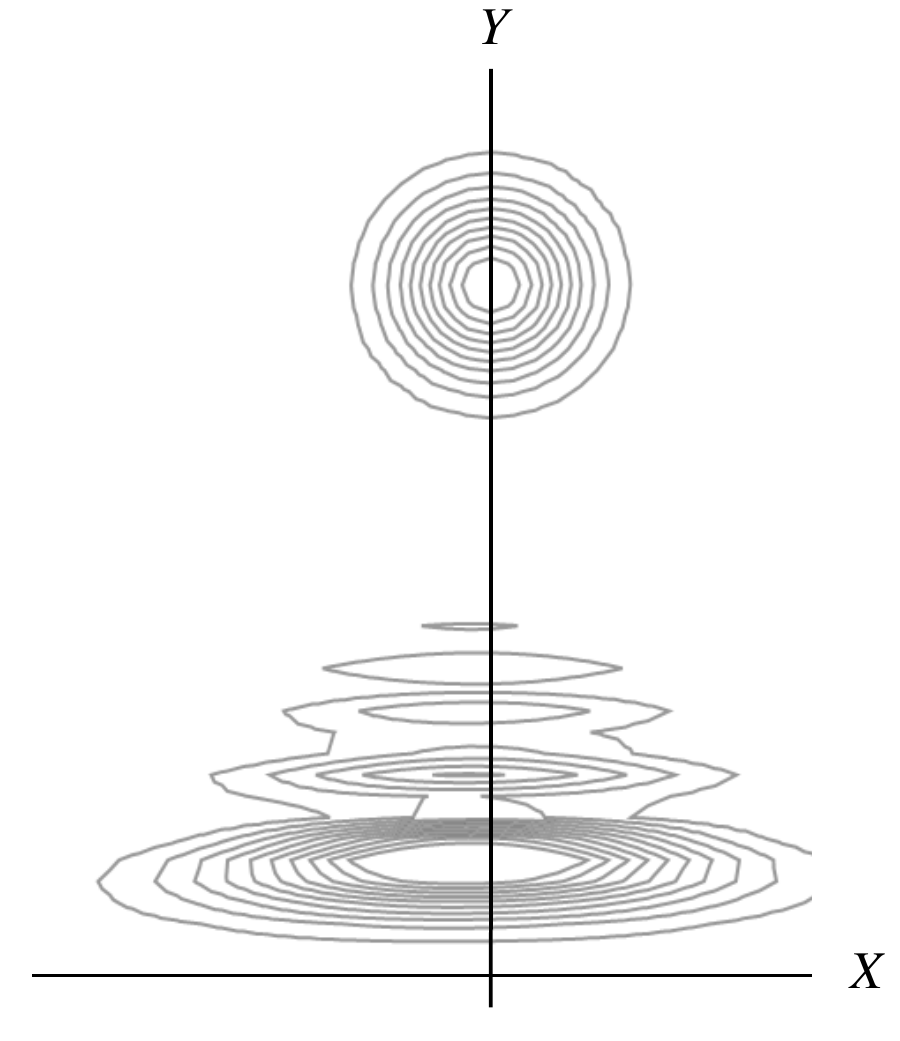}
\caption{Contour plots in the $XY$ plane corresponding to the case of the quasi-Bloch oscillations shown in Fig.\;\ref{fig5}.}
\label{fig6}
\end{figure}

In Fig.\;\ref{fig5}(a) we show the numerical simulations of the mean-value positions corresponding to an initial Gaussian wavepacket with conditions $(x_0,y_0,z_0)=(0,32,0)$ (in units of the lattice constant $a$) and $(k_{x0},k_{y0},k_{z0})=(0,0,0)$ (in units of $1/a$); the donnor impurity acting as center of the Coulomb attraction is located at $(x_0,y_0,z_0)=(0,-120,0)$ and the intensity of the field is $V_1=307200$ (adimensional units) in (\ref{1}). The motion of the particle is an approximate Bloch oscillation along the $Y$ axis with negligible values of $z_x(t)$ and $z_z(t)$ as compared to $z_y(t)$. The corresponding 2-dimensional simulation for $z_y(t)=\sum_{n,m} am|C_{nm}|^2$ is shown in Fig.\;\ref{fig5}(b). Thus, we confirm that the symmetrical dispersion of the wavepacket along the $Z$ axis does not affect the quasi-Bloch oscillation effect in the $XY$ plane (with dominant motion along the $Y$ axis), consistently with the semiclassical prediction about the 2-dimensional dynamics for the Kepler problem, being the symmetric dispersion of the wavepacket along the $Z$ axis an independent phenomenon which does not affect the evolution of $z_x$ and $z_y$. Further numerical confirmation that the quasi Bloch oscillation is well represented in the $XY$ plane is shown in Fig.\;\ref{fig6} for the countour plot of the probability density $|C_{nm}|^2$. In fact, the value of $z_x(t)$ at 100 \% of its time evolution is negative, as can be seen in Fig.\;\ref{fig5}(a).

\begin{figure}[H]
\centering
\includegraphics*[scale=0.40]{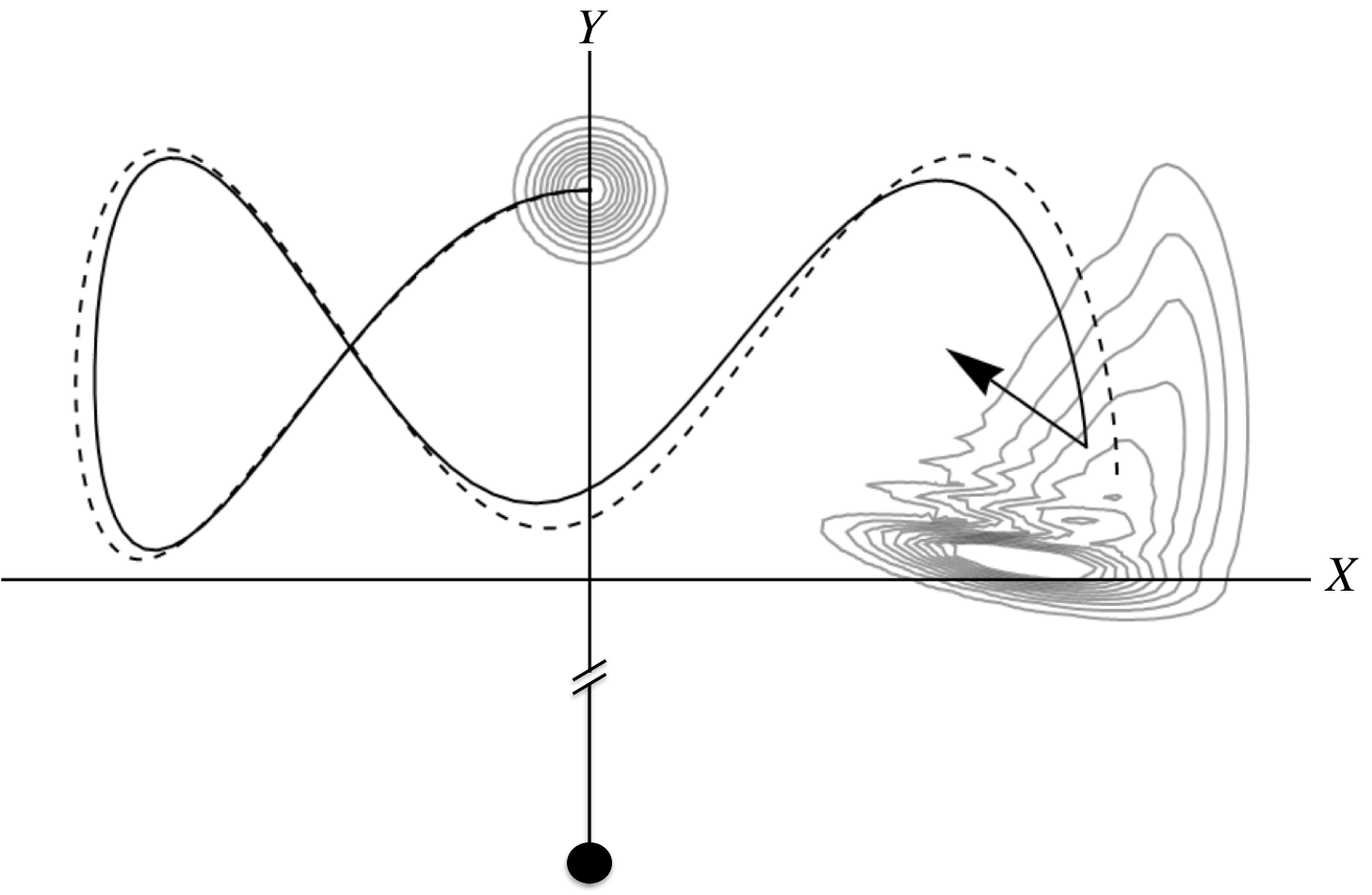}
\caption{Contour plots in the $XY$ plane of the probability density $|C_{nm}|^2$ of the initial (Gaussian) and final wavepackets. The initial conditions are $(x_0,y_0)=(0,32)$ (in units of the lattice constant $a$) and $(k_{x0},k_{y0})=(-1,0)$ (in units of $1/a$); the donnor impurity acting as center of the Coulomb attraction is located at $(0,-120)$ and the intensity of the field is $V_1=307200$ (adimensional units) in (\ref{1}). The solid-line trajectory of the mean-value position $\textbf{z}=(z_x,z_y)$ ends up at the position $(40.42,10.23)$. The arrow indicates the skewness vector $\textbf{s}=(-10.50,9.39)$ of the final wavepacket relative to its mean position. The dashed line corresponds to the semiclassical trajectory of the position $\textbf{r}=(x,y)$.}
\label{fig7}
\end{figure}

By changing the initial momentum from $k_{x0}=0$ (quasi-Bloch oscillation) to $k_{x0}=-1$, the resulting trajectory of the mean-value position $\textbf{z}=(z_x,z_y)$ is shown in Fig.\;\ref{fig7} (solid line); the contour plots show the probability density $|C_{nm}|^2$ of the initial Gaussian wavepacket and the final wavepaket with a skewness vector $\textbf{s}=(s_x,s_y)$, which indicates the asymmetry due to its deformation over time. The dashed line corresponds to the semiclassical trajectory of the position $\textbf{r}=(x,y)$ according to the Hamilton equations (\ref{2}), (\ref{3}), (\ref{5}), (\ref{6}).

\begin{figure}[H]
\centering
\includegraphics*[scale=0.5]{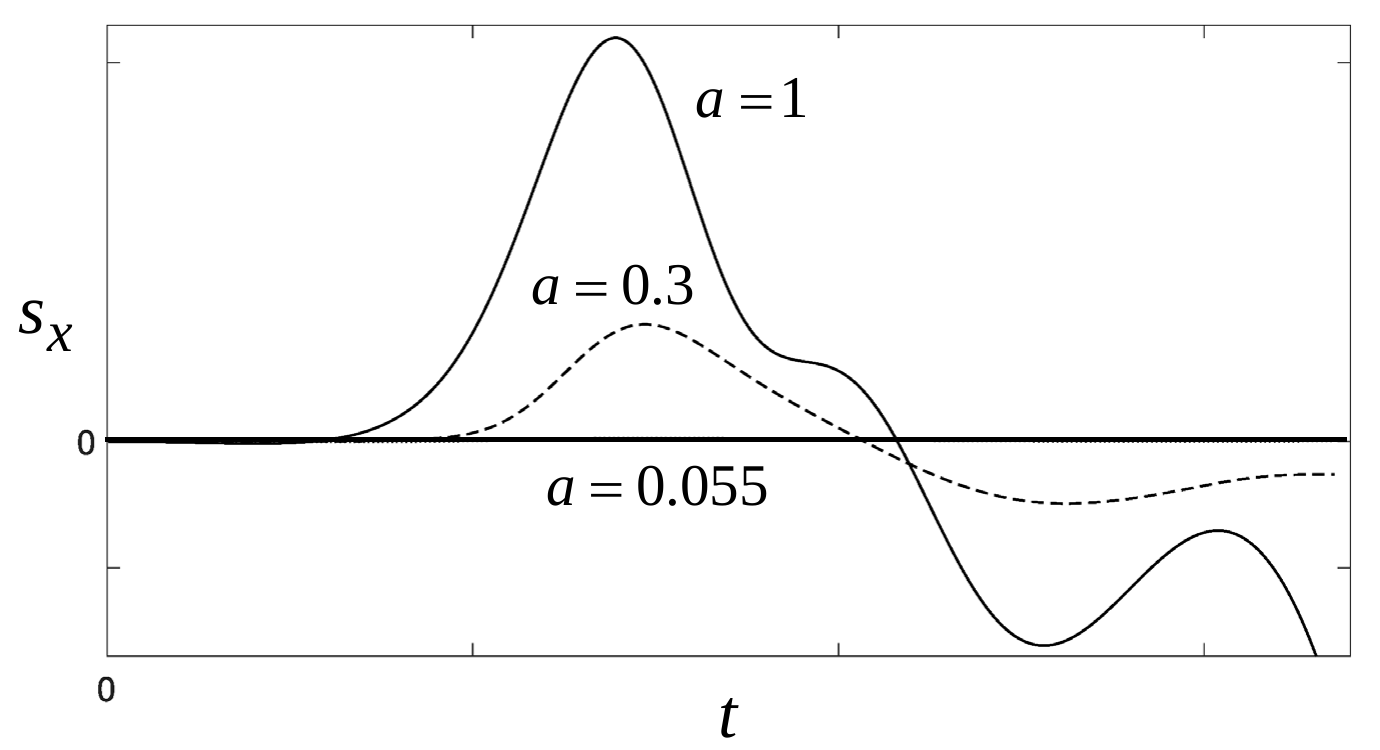}
\caption{Numerical evidence of $s_x(t) \rightarrow 0$ in the continuum limit for a particle moving under a Coulomb potential in a 1-dimensional lattice.}
\label{fig8}
\end{figure}

Next, we calculate the time evolution of the quantum angular momentum of the wavepacket approximated as $L_q=|\textbf{z}\times \textbf{k}|=z_x k_y-z_y k_x$, where the mean value of the position $\textbf{z}$ corresponds to the solid line in Fig.\;\ref{fig7}, and we compare it to the semiclassical angular momentum $L_c=|\textbf{r}\times \textbf{k}|=x k_y-y k_x$ where the position $\textbf{r}$ corresponds to the dashed line in Fig.\;\ref{fig7}. In both cases, $\textbf{z}$ and $\textbf{r}$ are measured relative to the impurity position which is in this case $\textbf{r}=(0,-120)$. The approximation referred to above for the quantum $L_q$ is that we use the same semiclassical momentum $\textbf{k}=(k_x,k_y)$ as for $L_c$. Such an approximation could be justified upon the following basis: it is known that in the continuum limit and under specific conditions \cite{Mermin 1976}, the quantum and semiclassical positions approximate very well for any time, $\textbf{z}(t) \cong \textbf{r}(t)$. In the lattice, however, a deviation should be introduced as $\textbf{z}=\textbf{r}+\alpha \textbf{s}$ such that $\textbf{s}\rightarrow 0$ in the continuum limit and $\alpha$ being a dimensionless scale factor. It happens that we already know of such a quantity: the modified ``skewness'' defined in this work as $\textbf{s}=(s_x,s_y)$ with $s_x^3\equiv\sum_{n,m}(an-z_x)^3|C_{nm}|^2$ and $s_y^3\equiv\sum_{n,m}(an-z_y)^3|C_{nm}|^2$. In Fig.\;\ref{fig8} a simulation of $s_x$ as $a \rightarrow 0$ was carried out for a 0hopping electron in a 1-dimensional lattice under the Coulomb field of a donor impurity located at the origin \cite{foot1}; we can verify that in fact $s_x \rightarrow 0$. We may reasonably suppose that in the 2-dimensional generalization of this problem, the Kepler problem in the lattice, it also occurs that $s_y \rightarrow 0$\cite{foot2}. Now, we propose in this work that $L_q=L_c+\alpha S$ with $L_q=|\textbf{z} \times \textbf{k}|$, $L_c=|\textbf{r} \times \textbf{k}|$ and $S=|\textbf{s} \times \textbf{k}|$ ($\textbf{z}$, $\textbf{r}$, $\textbf{s}$ and $\textbf{k}$ are coplanar vectors). The quantity $S$ represents an intrinsic angular momentum of the wavepacket due to its deformation in the lattice during its time evolution. In the limit of the continuum, $S \rightarrow 0$ and the quantum and semiclassical angular momenta coincide, $L_q=L_c$, as expected.

\begin{figure}[H]
\centering
\includegraphics*[scale=0.4]{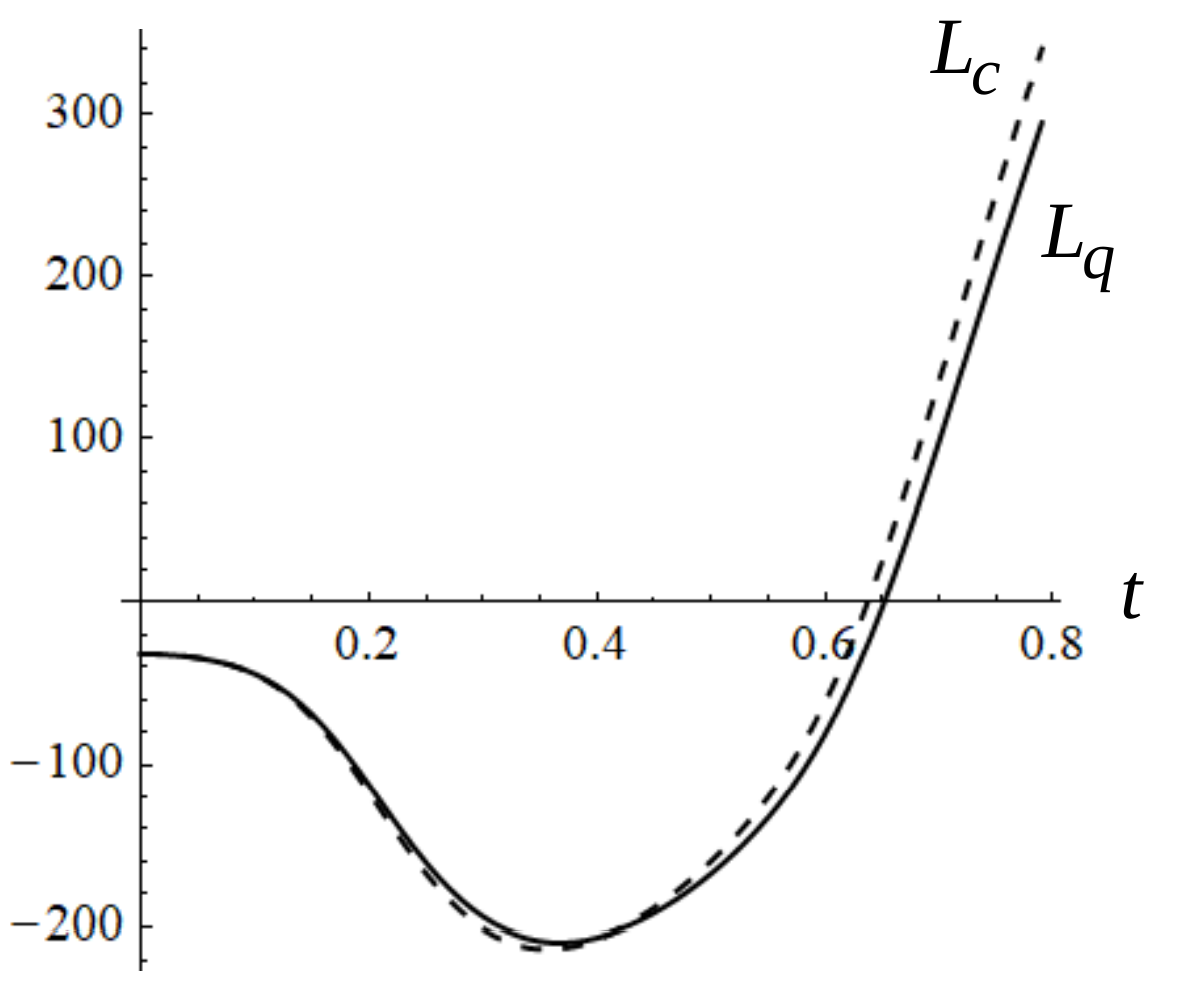}
\caption{Time evolution of the angular momenta corresponding to the case of Fig.\;\ref{fig7}: quantum $L_q(t)$ (solid line) and semiclassical $L_c(t)$ (dashed line).}
\label{fig9}
\end{figure}

\begin{figure}[H]
\centering
\includegraphics*[scale=0.4]{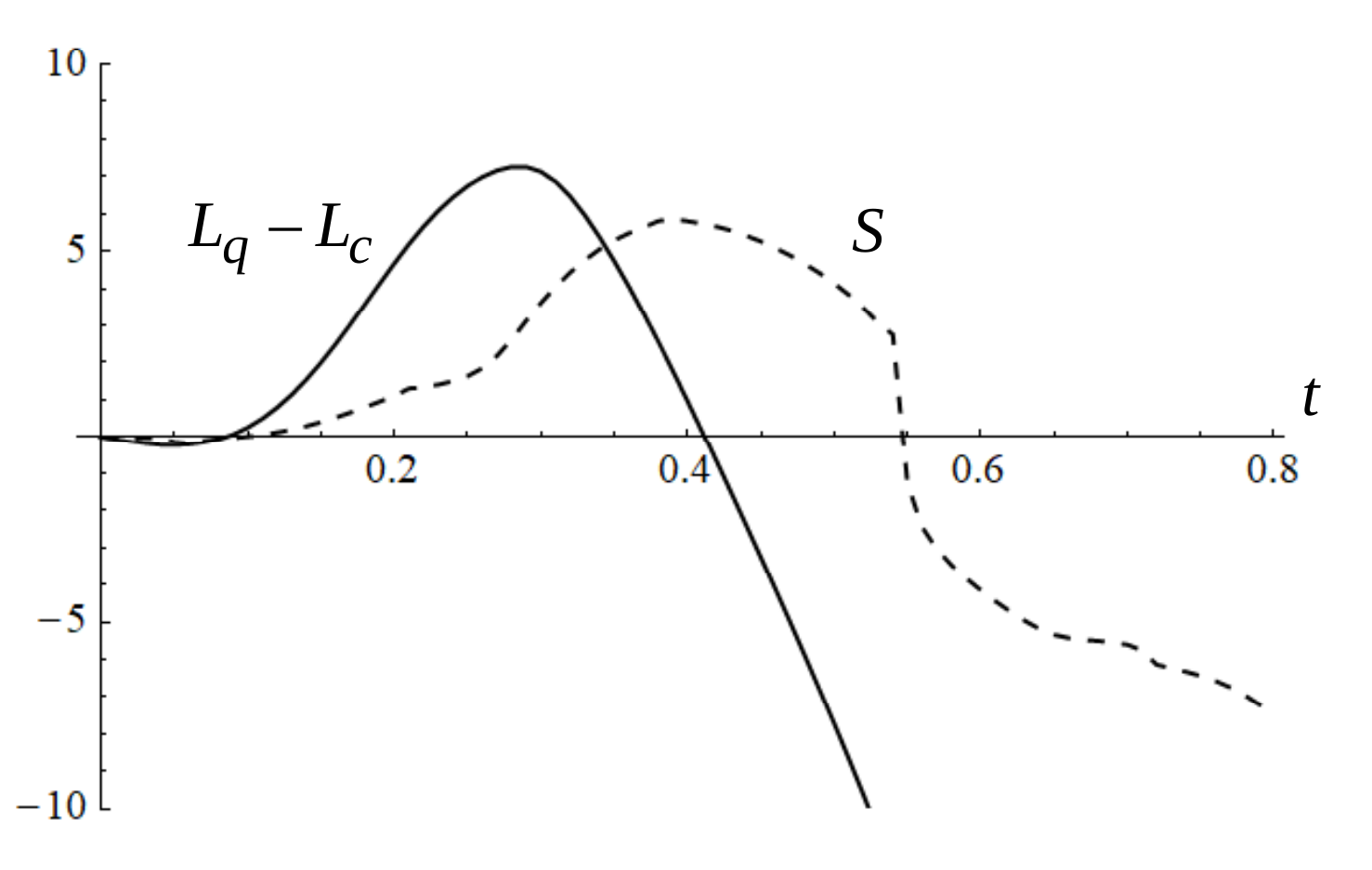}
\caption{Time evolution of the momenta difference $L_q(t)-L_c(t)$ corresponding to Fig.\;\ref{fig9} and the intrinsec angular momentum $S(t)$. The qualitative features of $L_q(t)-L_c(t)$ are roughly represented by $S(t)$, specifically: their change of sign at $t \cong 0.1$ and in the interval $0.4<t<0.6$.}
\label{fig10}
\end{figure}

In Fig.\;\ref{fig9} we show the numerical values of $L_q(t)$ (solid line) and $L_c(t)$ (dashed line). We can see that there is a quite good approximation between them. In order to quantify such an approximation we calculate their difference $L_q-L_c$ and compare it to the value of the intrinsic angular momentum $S$ (Fig.\;\ref{fig10}), where an arbitrary scale factor $\alpha$ has been chosen for the purpose of a better visual comparison. It is to be mentioned that other odd higher order statistical moments beyond the 3th (skewness) and usually refered to as ``hyperskewness" could also be used to refine our results. This task is left as an interesting issue for future work.

\section{Conclusions}

We have studied the motion of a charged particle (electron) in a rectangular lattice in the presence of the Coulomb electric field of another fixed charged particle (donor impurity). Although the field is 3-dimensional, we demonstrated that the motion will occur in a plane. The methods used in this study were the semiclassical model (Hamilton equations) and the quantum formalism; for the latter we invoked the tight-binding model for the Hamiltonian and the pseudo-spectral algorithm for the time integration of the Schr\"{o}dinger equation. The total energy is conserved but the angular momentum is not since the lattice discrete subspace has not rotational symmetry. We propose that these subjects are ideally suited for their pedagogical adaptation in undergraduate courses as classical mechanics, quantum mechanics and solid state.

Specifically, we report the following results:

i) We have reproduced a relevant result reported by Cort\'es \cite{Cortes2008} about the trajectory of a particle in an attractive central field. In \cite{Cortes2008} the author invokes different masses in the $X$ and $Y$ axes for the same particle. We have shown that this can be understood in terms of the continuum limit of a rectangular lattice.

ii) We have obtained the precession effect of a trajectory in a square lattice as a result of the non-conservation of the angular momentum. In the limit of the continuum we obtained a closed elliptical orbit (the precession disappears) and the angular momentum is conserved.

iii) We have applied the quantum formalism in 3 dimensions and integrated numerically the Schr\"{o}dinger equation. We have shown that the mean value of the particle's position remains in the lattice plane while the dispersion of the wavepacket in the normal direction does not affect the mean value of the position in the plane.

iv) We have used the quantum WKB approximation consistent with the semiclassical method to obtain the mean value of the angular momentum $L_q$. We have compared it with its equivalent semiclassical angular momentum $L_c$. We have proposed that their numerical difference can be attributed to a ``intrinsec angular momentum" $S$ of the wavepacket due to its asymmetrical deformation in the lattice: $L_q=L_c+\alpha S$, where $S$ is proportional to the skewness of a distribution (3th statistical moment). In the continuum limit ($a \rightarrow 0$), $S \rightarrow 0$ and $L_q \rightarrow L_c$.


\begin{thebibliography}{00}
\bibitem{foot2}
The numerical procedure to verify this assertion on the basis of the algorithm given in (\ref{11a}) is too demanding for a computer since the limit of the continuum $a \rightarrow 0$ implies the discretization of the lattice in a greater number of cells. For the numerical results presented in this work we have taken a region of $127 \times 127$ cells where a typical wavepacket evolution lasts around 30 minutes. The further segmentation $a\rightarrow a/2$ would imply a 4-fold time increment, and so on.
\end{thebibliography}
\end{document}